# Microstructured tunable two dimensional potential modulation in organic heterostructure field effect transistors

F.V. Di Girolamo

Consiglio Nazionale delle Ricerche (CNR), Italy

In this paper, two dimensional modulation of the potential in sexithiophene (T6) / N,N-bis(*n*-octyl)-dicyanoperylenediimide (PDI-8CN$_2$) heterojunction field effect transistors due to the specific microstructure at the interface is used to explain the negative transconductance effect (NTC) experienced in sexithiophene (T6) / N,N-bis(*n*-octyl)-dicyanoperylenediimide (PDI-8CN$_2$) heterojunction field effect transistors. The NTC effect has been experienced in tunnel devices, such as the tunnel diode, the resonant tunneling field effect transistors (RT-FETs), resonant tunneling double barrier devices. In grid-gate modulation-doped field effect transistors, instead, a periodic potential barriers in the direction of the transport of charges was used to explain the negative transconductance (NDR). Since in T6 / PDI-8CN$_2$ heterojunction field effect transistors the NTC effect is irrespective of the order of the semiconductor layer and since the modulation of the transport properties is deeply influenced by the island dimension of the semiconductor layer, we argue that the origin of the NTC effect resides in the achievement of a specific microstructure of the heterostructure in the charge transport plane.

The semiconductor device field is almost totally based on junctions of different kinds [1]. Several junction devices exhibit a decrease and nullification of the current for increasing voltages, the so called negative transconductance effect (NTC) [1] : the Esaki diode [2], the resonant tunneling double barrier devices [3], and the resonant tunneling field effect transistors (RT-FETs) [4]. NTC was also experienced in grid-gate modulation-doped field effect transistors [5].

The realization of analogous devices with organic materials consequently requires the understanding and the control of the microstructure and of the electronic structure [6] [7] at the junction between two organic semiconductors.

Resonant tunneling diodes have been realized with a superlattice of amorphous layers of [1 , 3 , 5 – tris – (3 – methylphenylphenylamino) triphenylamine (*m*-MTDATA) / 4 , 7 diphenyl – 1 , 10 - phenanthroline (Bphen)]$_4$ [8] exhibiting in some cases a negative differential resistance (NDR) for low voltages. A high peak to valley current ratio allowed the demonstration of a basic logic circuit operation in poly[2-methoxy-5-(2'-ethyl-hexyloxy)-1,4-phenylenevinylene] (MEH-PPV) polymer tunnel diodes (PTD) using a thin $TiO_2$ tunneling layer; the NDR was explained by tunneling through defect states present in the $TiO_2$ [9].

NTC effect was also experienced in sexithiophene (T6) / N,N-bis(*n*-octyl)-dicyanoperylenediimide (PDI-8CN$_2$) heterojunction field effect transistors [10]. In those devices, the effect was explained in terms of gate-voltage-tunable recombination phenomena occurring in the hole/electron interface, assuming the formation of an accumulation junction at the interface and that the heterojunction could be modeled as formed by perfect parallel layers (see fig. 1). For the case of T6 as lower layer, it was supposed that when the thickness of the T6 layer is smaller than the hole accumulation length, the electric field due to the application of a potential to the gate penetrates inside the T6 up to the interface with PDI8-CN$_2$ and hole accumulation extends throughout the whole T6 layer. Consequently, both voltage profile and hole density redistribute also in the region close to the T6/PDI-8CN$_2$ interface which causes a flattening of the molecular bending of T6 levels at the heterojunction interface.

In this study, an alternative explanation for the NTC effect in T6 / PDI-8CN$_2$ heterostructure field effect transistors has been presented, which suggests an alternative route for the realization of organic junctions and field effect devices. Based on recent in situ x-ray (XPS) and ultraviolet (UPS) photoemission spectroscopy measurements [11], the explanation takes into account the actual tridimensional microstructure of the heterojunction interface due to the island growth of the organic semiconductors; moreover, it can be immediately extended to the case of PDI-8CN$_2$ as lower layer, which was not clearly explained until now.

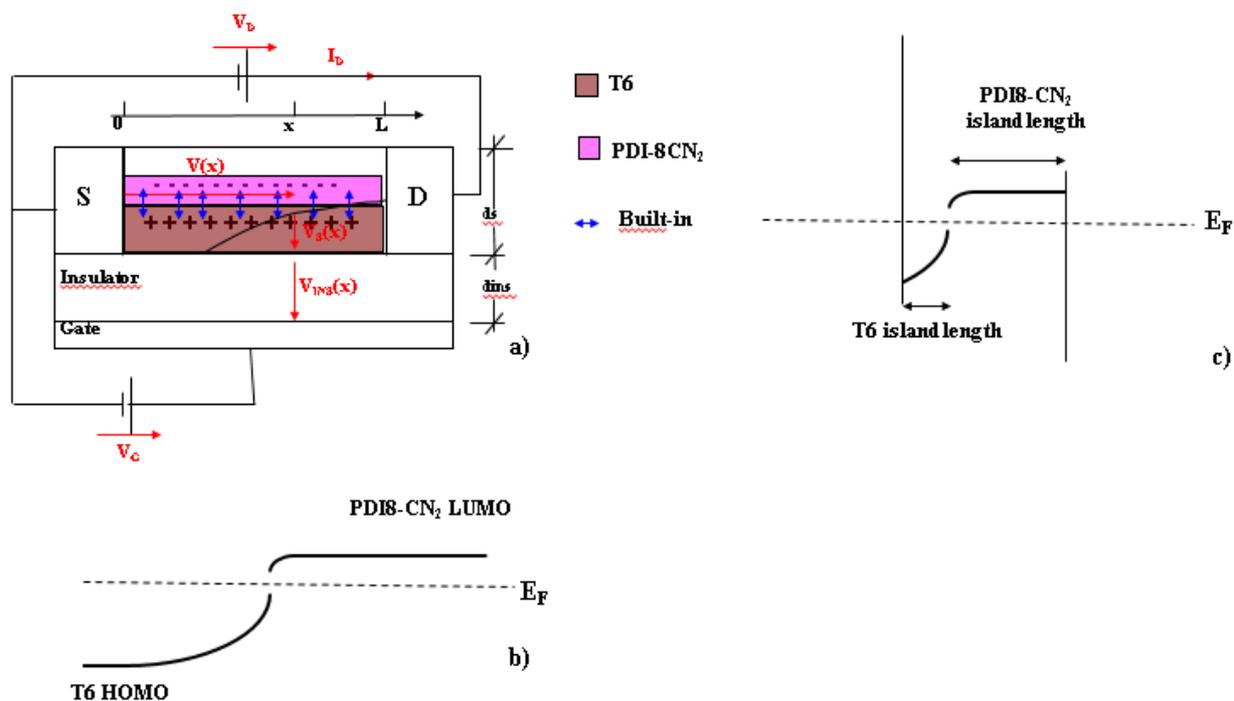

Fig.1. a) schematical representation of the T6 / PDI8-CN$_2$ (T6 as lower layer) field effect transistor also illustrating the accumulation junction; b) band structure of a T6 / PDI8-CN$_2$ interface; c) band structure of a T6 / PDI8-CN$_2$ interface supposing finite islands.

In order to correctly discuss the energetics of the T6 / PDI-8CN$_2$ heterostructure, the schematic energy level diagrams of T6 and PDI-8CN$_2$ on heated Au is reported as deduced from, respectively, [12] [13] and [11] (fig. 2 a and b). The choice of heated Au (i.e. contaminated Au, not sputter cleaned gold ) is motivated by the need for discuss the most realistic situation as possible, closer to the one of real devices [10], in which the substrates are only heated and not sputter cleaned before the deposition of the organic layer. As widely discussed in [12] [13] and [11], the energy level of organic semiconductors on contaminated substrates highly differs from the one on sputter cleaned ones. In particular, it is possible to note that the T6 HOMO level and the PDI-8CN$_2$ LUMO level are closer to the Fermi level of the system. PDI-8CN$_2$ LUMO level, in particular, results almost superimposed to the Fermi level (as also found in [14]) highly favouring the injection of electrons in PDI-8CN$_2$. This suggests that at room temperature and equilibrium conditions electrons are naturally accumulated in PDI-8CN$_2$, explaining the outstanding transport properties of this semiconductor, which exhibit a negative threshold voltage in field effect transistors [15] [16].

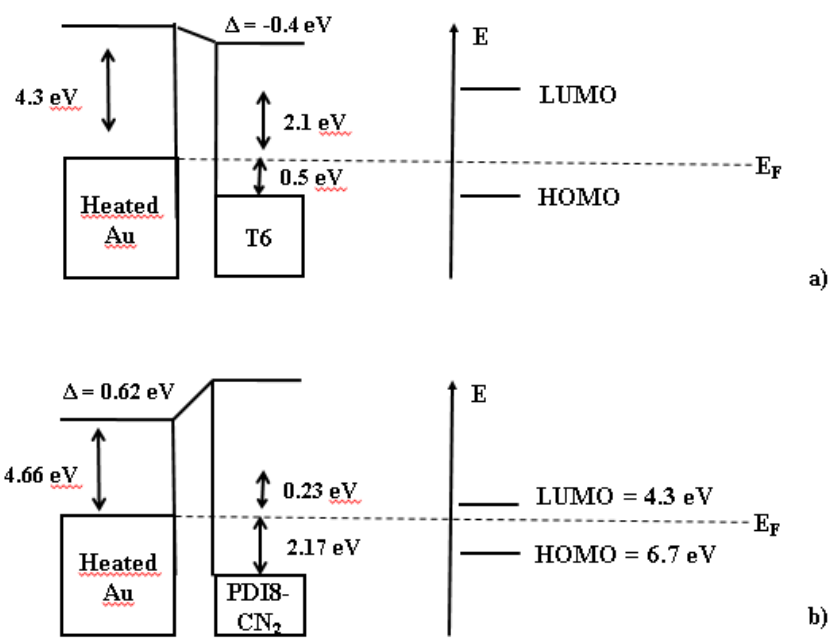

Fig. 2 : a) energy levels of T6 on heated (i.e. not sputtered cleaned ) Au (from [12] and [13]); b) energy levels of PDI-8CN$_2$ on heated (i.e. not sputtered cleaned ) Au (from [11]).

As already mentioned, in T6 / PDI-8CN$_2$ heterostructure field effect transistors an accumulation junction is formed at the interface between the two semiconductors (fig. 1 a) [10]. Holes are accumulated in the p-type semiconductor and electron in the n-type one, resulting in an energy level diagram with an upward ending of the T6 HOMO level and a downward one of the PDI-8CN$_2$ LUMO (fig. 1 b ) [10]. The consequent formation of a built – in potential at the interface between in T6 and PDI8-CN$_2$ was actually demonstrated by transport measurements, Second Harmonic Generation (SHG) [10], ex- situ Ultraviolet Photoemission Spectroscopy (UPS) [14] and in situ X-ray photoemission spectroscopy (XPS) and UPS measurements [11].

The in situ XPS and UPS measurements have clearly demonstrated, in agreement with the ex-situ UPS measurements, that the accumulation and band bending phenomena are more evident from the T6 side (see fig. 2 (b) : T6 is doped by PDI8-CN$_2$ [14] n.d.r.), while PDI8-CN$_2$ seems not to be dramatically influenced by the presence of T6 but more by the substrate. This is not surprising, if we take into account that, as previously discussed, in PDI8-CN$_2$ the LUMO level is already closer to the Fermi level.

In addition, the in situ XPS and UPS measurements [11] have demonstrated that the energy levels are not irrespective of the order of the organic layers. Since the NTC effect appears independently of the order of the layers, this suggests that the band bending alone cannot be used to produce a complete explanation of the NTC phenomenon and that another factor has to be taken into account, such as the actual microstructure of the junction. If the organic layers are not infinite, the energy bands do not go to the equilibrium: the band diagram reported in fig.1 b) has to be modified like the one fig. 1 c) and the T6 HOMO level results consequently lower than in the previous case.

Organic materials, T6 and PDI8-CN$_2$ in particular, are characterized by an island -like growth morphology: if T6 thickness is lower than 2.5 nm the film is not conductive [17]; PDI8-CN$_2$ is instead characterized by an elongated island morphology which strongly depends on the substrate [18]. This suggests that the actual microstructure of T6 / PDI8-CN$_2$ heterostructures is formed by alternating islands of T6 and PDI8-CN$_2$, as depicted in fig. 3a, where a schematic representation of the whole device is reported, without any drain and gate voltage applied. It is worth to mention that, contrary to the hypothesis of flat parallel layers, in this case vertical, periodic T6 / PDI8-CN$_2$ interfaces with the relative built-in potential have to be considered (i.e. fig. 3 a instead of fig. 1 a).

A schematic of the equilibrium energy levels of the alternating T6 and PDI8-CN$_2$ islands is reported in fig. 3 b and highlighted in fig. 3 c, as obtained by sticking the energy levels reported in fig. 1 c for finite T6 / PDI8-CN$_2$ heterostructures; a periodic variation of the potential is obtained. The device structure can be considered as several p-n junctions in series, in which an accumulation instead than a depletion junction is created at the interface and the p [n]-type valence [conduction] band is replaced by the T6 HOMO [PDI8-CN$_2$ LUMO] level.

Let us now consider the application of a drain voltage. Since several junctions in series are present, some of them will experience a forward bias, others a reverse bias. In depletion p-n junctions, the application of a forward bias causes the holes to be pushed toward the interface, resulting in a decrease of the depletion width; a reverse bias, instead, causes an increase of the depletion width [1]. The Fermi level is no more the same throughout the whole system, since equilibrium conditions are no more fulfilled.

For accumulation junctions, pushing holes toward the interface results in an increase of the accumulation width and a decrease in the opposite case, resulting the energy levels reported in fig. 4 a. The energy levels of fig. 1 b and c transform into the levels of fig. 4 b (in case of forward bias) and 4 c (in case of reverse bias). If several junctions in series are considered, the resulting energy diagram (n.d.r. except from the band bending which would complicate too much the picture) will be the one reported in fig. 4 d: the alternating and decreasing energy levels somehow resemble the enegy band diagrams of resonant tunneling diodes under bias [1], which exhibits the negative differential resistance (NDR), except for the absence of an insulating layer between two semiconducting ones and for the fact that energy levels are not quantized.

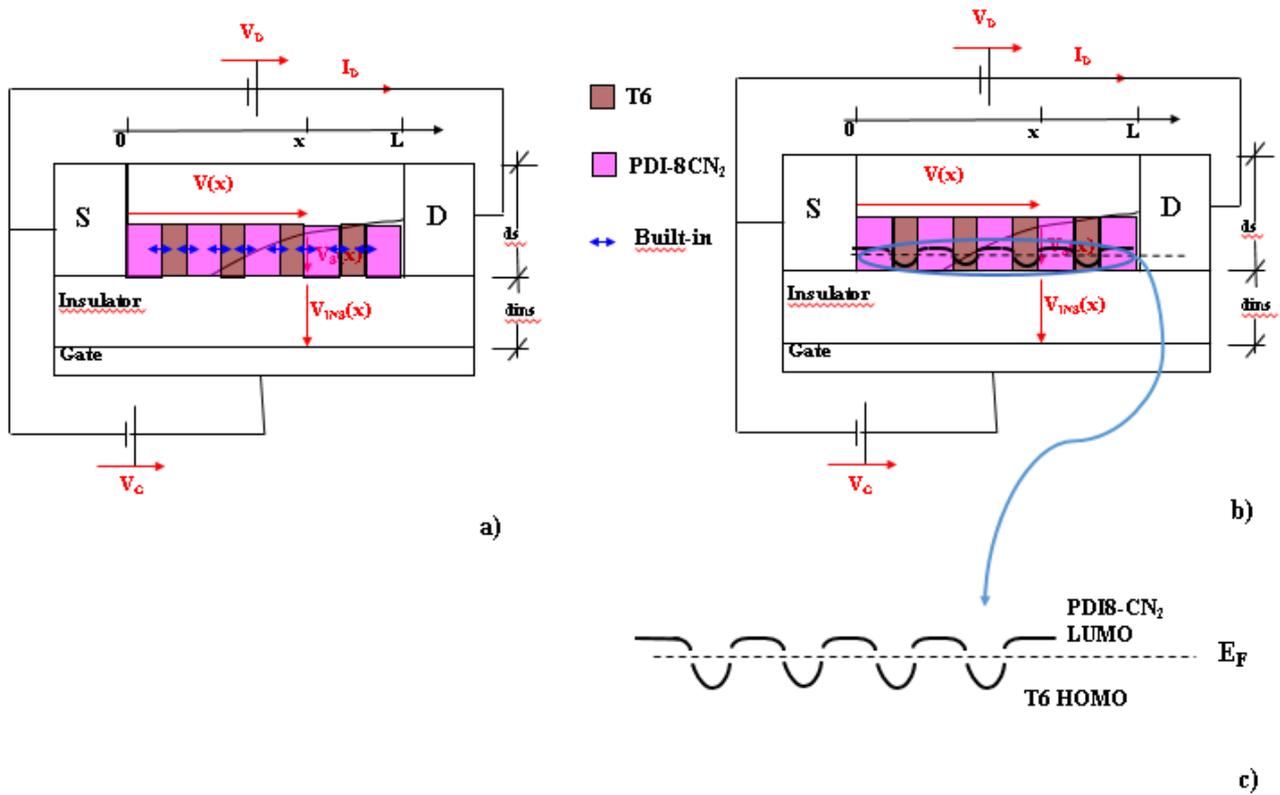

Fig. 3: schematical representation of the T6 / PDI8-CN$_2$ (T6 as lower layer) field effect transistor illustrating a) the interface microstructure with the built-in potential and b) the periodic band structure resulting from the alternating islands. c) The periodic microstructure highlighted.

The last condition to be applied to the system in order to describe the operation of T6 / PDI-8CN$_2$ heterostructure field effect transistors, is the application of a negative gate bias. Without further complicating the reasoning, we can simply say that, at sufficiently negative voltages, electrons are completely depleted in PDI-8CN$_2$, which actually becomes an insulator. Moreover, since holes in T6 are mostly accumulated because of the doping effect at the heterointerface, depleting PDI-8CN$_2$ means also strongly reducing the density of holes in T6.

The situation, at this point, than completely resembles the one of a resonant tunneling diode, except for the fact that the levels are not quantized. If the PDI-8CN$_2$ islands are sufficiently extended (i.e. when the T6 layer is really low and T6 islands are not coalesced) the probability of hole tunneling through the PDI-8CN$_2$ layer is too low to provide a sufficiently high current. The tunneling probability is further decreased, especially in smaller islands, by other two factors: first of all, T6 HOMO is already lower at equilibrium, as discussed above (fig. 1 c); secondly, the hole density is dramatically reduced as consequence of PDI-8CN$_2$ depletion (n.d.r. it is something similar but opposite to the case of [19]; in our case it is like "removing" the doping).

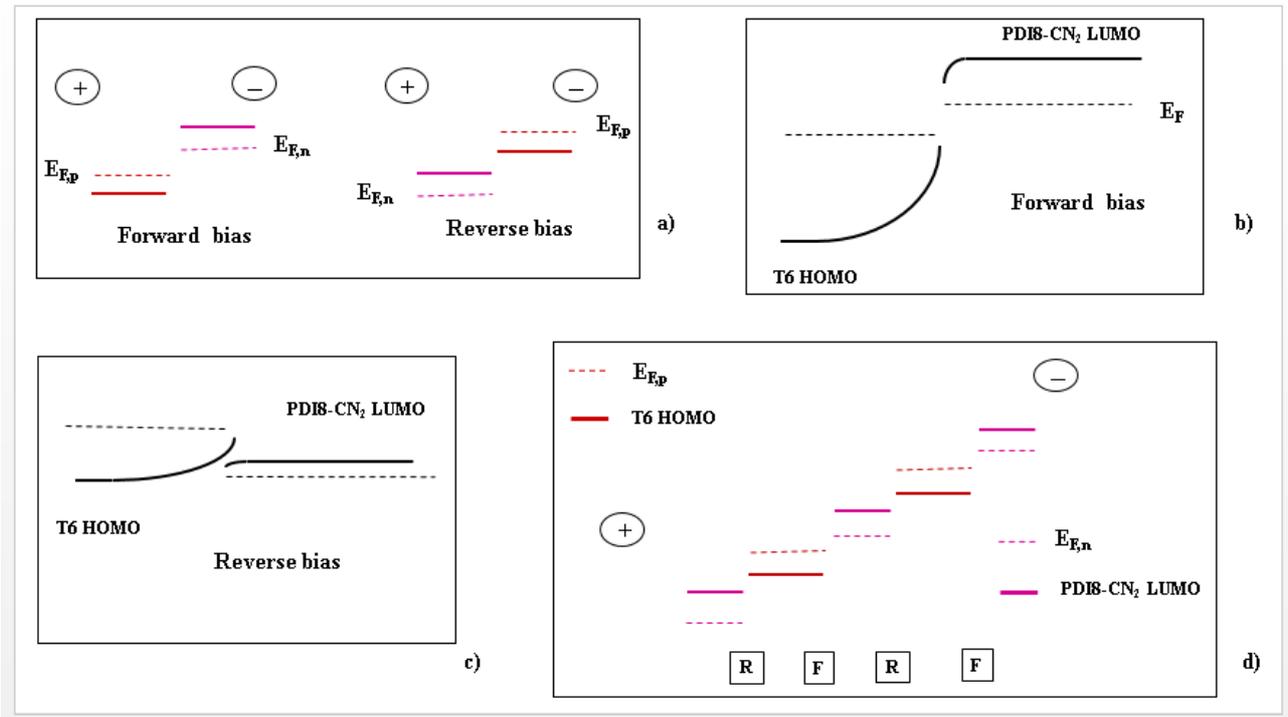

Fig. 4 : a) schematical energy level for a single T6 / PDI8-CN$_2$ accumulation junction in case of forward (left)and reverse (right) bias; b) band structure in case of forward bias; c ) band structure in case of forward bias, d)  schematical energy level for a series of T6 / PDI8-CN$_2$ accumulation junctions biased.

Fig. 5 summarizes the different regions of the device transfer curve. In the region I ( $V_{DS}$>0, $V_{DS}$ > PDI-8CN$_2$ threshold voltage, $V_{DS}$ > T6 threshold voltage) electrons are accumulated in PDI-8CN$_2$ and T6 is depleted. In the region II ($V_{DS}$ > PDI-8CN$_2$ threshold voltage, $V_{DS}$ < T6 threshold voltage) charges are accumulated in both the semiconductors: it is worth to notice that charges are accumulated both for the application of the gate and for the band bending at the heterointerface; this is especially true for T6. In the region III charges in PDI-8CN$_2$ are depleting. In the region IV the holes tunneling current through "insulating" PDI-8CN$_2$  islands is sufficiently high.

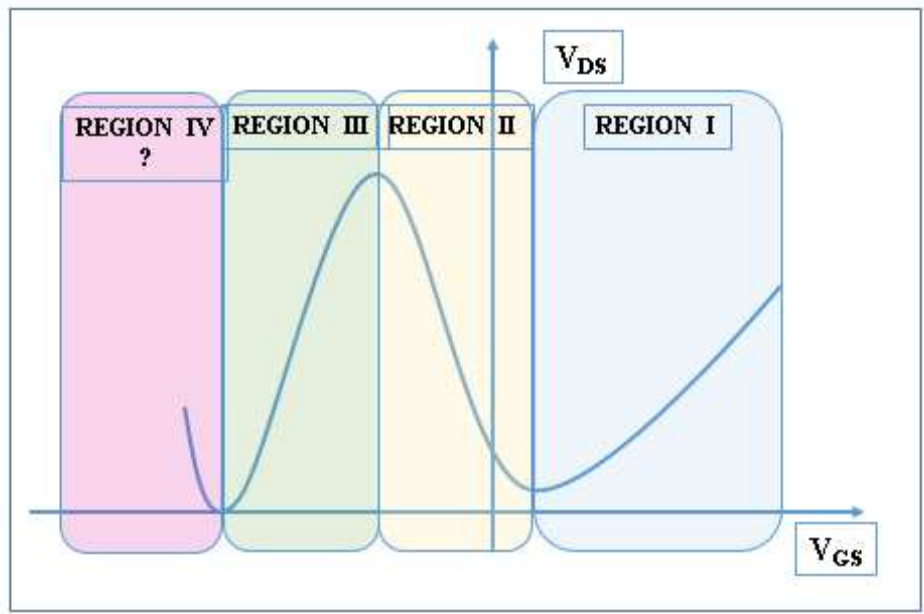

Fig. 5. Typical transfer curve for a T6 / PDI-8CN$_2$ heterostructure field effect transistor illustrating the different regions of conduction: Region I PDI-8CN$_2$ only ; Region II T6 and PDI-8CN$_2$ ; Region III PDI-8CN$_2$ depleting; Region IV (hypothetical) T6 only tunneling.

When the T6 thickness increases, PDI-8CN$_2$ island dimension decreases, consequently increasing the tunneling probability; T6 islands start to form a continuous conductive layer which can sustain a current even without doping; T6 HOMO level for the islands also increases.

The explanation can easily extended to the case of as PDI-8CN$_2$ lower layer: basically, when a negative voltage is applied, PDI-8CN$_2$ depletes and the current is only sustained by the holes accumulated by the gate; when PDI-8CN$_2$ is fully depleted the probability of hole tunnel is also null. The reason why the NTC effect appears only for negative voltages may reside in the fact that, while PDI-8CN$_2$ is "naturally" doped, T6 is instead doped by PDI-8CN$_2$: if the density of charge accumulated by the gate voltage is too low or lower than the one accumulated at the heterojunction, than the current drops or eventually go to zero.

Nevertheless, the p – alternating microstructure and the insulating barrier provided by PDI-8CN$_2$ at highly negative gate voltages is also necessary to explain the NTC phenomenon: for T6 thicknesses higher than 2.5 nm the films are expected to sustain a current with or without PDI-8CN$_2$; if the current goes to zero another phenomenon has to be considered. Moreover, it cannot be a simple depletion like in [19] due to a shift of the threshold voltage: the increase of the current for negative voltages means that the T6 threshold voltage has been already reached, even if shifted by the doping. Finally, it is not possible to suppose that the current nullifies after the peak because of a shift of the threshold voltage due to PDI-8CN$_2$ depletion (i.e. due to the removal of the dopant), because it cannot shift the threshold voltage to more negative values than the T6 alone (which are supposed to be less, in modulus, than -30 V, even for low thickness, as suggested by the measurements in [10]).

Since current in organic thin film is normally considered as the result of the tunneling (i.e. through grain boundaries [20]), than a stronger hindrance to tunneling has to be taken into account.

It is worth to mention that in all our reasoning the probability of recombination at p / n interface has been always considered by tacit agreement negligible. In depletion p / n junctions the recombination is also excluded at the interface, but this is also the main reason why it has been also neglected in our reasoning. This work aims to explain the appearance of the NTC phenomenon in T6 / PDI-8CN$_2$ heterostructure field effect transistors. The NTC phenomenon results dependent from T6 thickness (for the reasons exposed above) and on the gate voltage. While the application of a drain voltage can increase or reduce the accumulation width (depending on the disposition of the layers respect to the applied bias), the gate voltage has always the effect to reduce the electrons accumulation width. Since the probability of recombination increases when the density of both charge carriers increases and since gate voltage decreases electron density, than gate voltage cannot increase recombination probability. Consequently, recombination cannot be the explanation of the current decrease.

As already discussed in [10], trapping can be also excluded as the origin of the phenomenon both for the absence of hysteresis and for the strong and regular dependence on the T6 thickness.

In conclusion, the NTC effect in T6 / PDI-8CN$_2$ heterostructure field effect transistors has been explained in terms of a periodic potential consequent to the specific microstructure achieved at the junction between the two semiconductors. The modulation of the potential due to the application of the gate causes PDI-8CN$_2$ charges to deplete and the probability of hole tunneling through PDI-8CN$_2$ "insulating" islands to drop. This consideration suggests that with a careful control of the microstructure and a careful analysis of the energy levels at the junction between organic semiconductors, novel functionalities and device architectures can be achieved.